\documentclass[journal]{IEEEtran}

\usepackage[pdftex]{graphicx}
\usepackage{color,soul}
\usepackage{amsbsy,amssymb,latexsym,amsmath,amsthm,yhmath}
\usepackage{multirow}
\usepackage{caption}
\usepackage{subcaption}
\usepackage{listings}
\usepackage[numbers,sort&compress]{natbib}

\usepackage{soul}
\usepackage{xcolor}

\begin{document}



\title{Stability and Performance Limits of Latency-Prone Distributed Feedback Controllers}


\author{Y. Zhao, N. Paine, K.S. Kim, and L. Sentis$^*$
\thanks{Y. Zhao, K.S. Kim and L. Sentis (* is the corresponding author) are with the Department of Mechanical Engineering, The University of Texas, Austin, TX 78712 USA (e-mail: yezhao@utexas,edu, kskim@utexas.edu, lsentis@austin.utexas.edu).}
\thanks{N. Paine is with the Department of Electrical and Computer Engineering, The University of Texas, Austin, TX 78712 USA (e-mail: npaine@utexas.edu).}
}

\maketitle

\begin{abstract}
Robotic control systems are increasingly relying on distributed feedback controllers to tackle complex sensing and decision problems such as those found in highly articulated human-centered robots. These demands come at the cost of a growing computational burden and, as a result, larger controller latencies. 
To maximize robustness to mechanical disturbances by maximizing control feedback gains, this paper emphasizes the necessity for compromise between high- and low-level feedback control effort in distributed controllers. Specifically, the effect of distributed impedance controllers is studied where damping feedback effort is executed in close proximity to the control plant and stiffness feedback effort is executed in a latency-prone centralized control process. A central observation is that the stability of high impedance distributed controllers is very sensitive to damping feedback delay but much less to stiffness feedback delay. This study pursues a detailed analysis of this observation that leads to a physical understanding of the disparity. Then a practical controller breakdown gain rule is derived to aim at enabling control designers to consider the benefits of implementing their control applications in a distributed fashion. These considerations are further validated through the analysis, simulation and experimental testing on high performance actuators and on an omnidirectional mobile base. 
\end{abstract}


\newtheorem{theorem}{Theorem}
\newtheorem*{theorem*}{Theorem}


\section{Introduction}
\label{sec:intro}
\begin{figure}
 \centering
   \includegraphics[width=\linewidth]{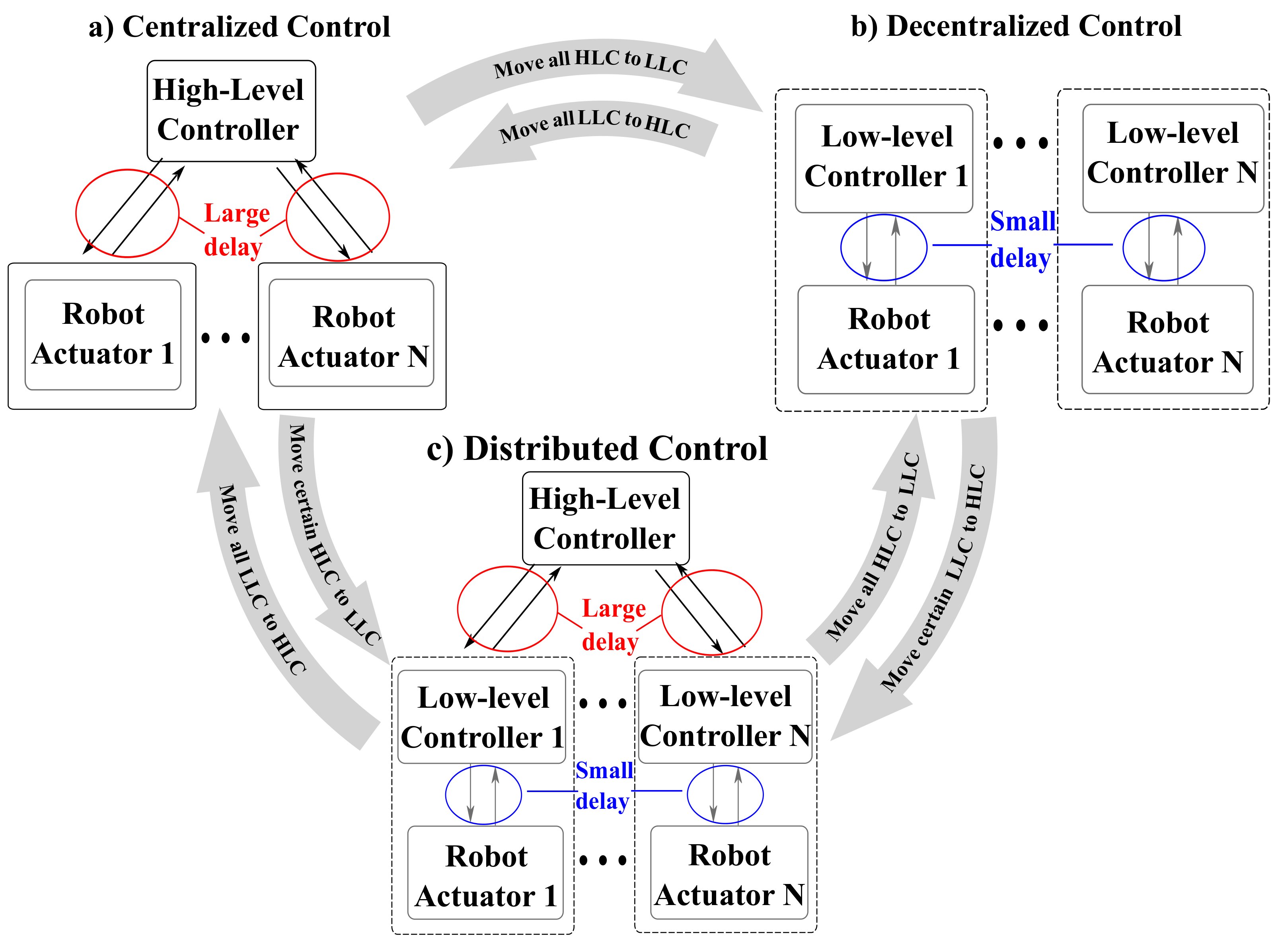}
 \caption{\textbf{Depiction of Various Control Architectures.} Many control systems today employ one of the control architectures above: a) Centralized control with only high-level feedback controllers (HLCs); b) Decentralized control with only low-level feedback controllers (LLCs); c) Distributed control with both HLCs and LLCs, which is the focus of this paper.}
 \label{fig:model}
\end{figure}

As a result of the increasing complexity of robotic control systems, such as human-centered robots \cite{sakagami2002intelligent, diftler2011robonaut}
and industrial surgical machines \cite{okamura2004methods}, new system architectures, especially distributed control architectures \cite{kim2005system, santos2006design}, 
are often being sought for communicating with and controlling the numerous device subsystems. Often, these distributed control architectures manifest themselves in a hierarchical control fashion where a centralized controller can delegate tasks to subordinate local controllers (Figure \ref{fig:model}). As it is known, communication between actuators and their low-level controllers can occur at high rates while communication between low- and high-level controllers occurs more slowly. 
The latter is further slowed down by the fact that centralized controllers tend to implement larger computational operations, for instance to compute system models or coordinate transformations online.

\subsection{Control architectures with feedback delays}
One concern is that feedback controllers with large delays, such as the centralized controllers mentioned above, are less stable than those with small delays, such as locally embedded controllers. Without the fast servo rates of embedded controllers, the gains in centralized controllers can only be raised to limited values, decreasing their robustness to external disturbances \cite{lu2014performance} and unmodelled dynamics \cite{martin1981continuous, gutman1979uncertain}.

As such, why not removing centralized controllers altogether and implementing all feedback processes at the low-level? Such operation might not always be possible. For instance, consider controlling the behavior of human-centered robots (i.e. highly articulated robots that interact with humans). Normally this operation is achieved by specifying the goals of some task frames such as the end effector coordinates. One established option is to create impedance controllers on those frames and transform the resulting control references to actuator commands via operational space transformations \cite{Khatib:87(2)}. Such a strategy requires the implementation of a centralized feedback controller which can utilize global sensing data, access the state of the entire system model, and compute the necessary models and transformations for control. Because of the aforementioned larger delays on high-level controllers, does this imply that high gain control cannot be achieved in some human centered robot control due to stability problems? It will be shown that this may not need to be the case. But for now, this delay issue is one of the reasons why various currently existing human-centered robots cannot achieve the same level of control accuracy that it is found in high performance industrial manipulators. More concretely, this study proposes a distributed impedance controller where only proportional (i.e., stiffness) position feedback is implemented in the high-level control process with slow servo updates. This process will experience the long latencies found in many modern centralized controllers of complex human-centered robots. At the same time, it contains global information of the model and the external sensors that can be used for operational space control. For stability reasons, our study proposes to implement the derivative (i.e., damping) position feedback part of the controller in low-level embedded actuator processes which can therefore achieve the desired high update rates.

\subsection{Analysis of sensitivity to delay}
To focus the study on the physical performance of the proposed distributed control approach, our study first focuses on a single actuator system with separate stiffness and damping servos and under multiple controller delays. Then the physical insights gained are used as a basis for achieving high impedance behaviors in single actuator systems and in an omnidirectional mobile base.
Let us pose some key questions regarding distributed stiffness-damping feedback controllers considered in this paper: (A) Does controller stability have different sensitivity to stiffness and damping feedback delays? (B) If that is the case, what are the physical reasons for such a difference? (C) In applications where load uncertainty exists, how robust is the distributed controller to these uncertainties?

To answer these questions, this paper studies the physical behavior of the proposed realtime distributed system using control analysis tools applied to the system's plant, including the phase margin stability criterion and time-based trajectory tracking response. Using these tools our study reveals that system stability and performance are much more sensitive to damping feedback delays than to stiffness feedback delays. 

\subsection{Benefits of the proposed distributed control architecture}
As it will be empirically demonstrated, the benefit of the proposed split control approach over a monolithic controller implemented at the high-level is to increase control stability due to the reduced damping feedback delay. As a direct result, closed-loop actuator impedance may be increased beyond the levels possible with a monolithic high-level impedance controller. This conclusion may be leveraged on many practical systems to improve disturbance rejection by increasing gains without compromising overall controller stability. As such, these findings are expected to be immediately useful on many complex mechatronic and control systems. 

To demonstrate the effectiveness of the proposed methods, this study implements two types of tests on a high performance actuator followed by experiments on a mobile base. First, a position step response is tested on an actuator under various combinations of stiffness and damping feedback delays. The experimental results show high correlation to their corresponding simulation results. Second, trajectory tracking performance of a complex joint trajectory with load uncertainty is tested both in simulation and in hardware on the same actuator. Third, the proposed distributed controllers is leveraged an implementation into an omnidirectional base. The results show a substantial increase in closed-loop impedance capabilities, which results in higher tracking accuracy with respect to the monolithic centralized controller counterpart approach.

\section{Related Work}
\label{sec:relatedwork}
Advances in distributed control technologies \cite{wang2011decentralized,santos2006design} have enabled the development of decentralized multi input and multi output systems such as humanoid systems and highly articulated robots 
\cite{sakagami2002intelligent,diftler2011robonaut}. However the effect on controller performance due the ever growing computational demand on feedback servos and latency-prone serial communications in human-centered robots have been largely overlooked on these studies. 

Distributed control architectures in robotics combine centralized processes with self-contained control units in close proximity to actuators and sensors. Two inherent advantages of having low level processes are: 1) to increase computational resources by offloading the workload of the central computer onto distributed processors \cite{kim2005system}
and 2) to increase control system stability and tracking performance due to reduced feedback delays. This study focuses on the latter advantage, i.e., how a distributed controller improves control system stability and performance over monolithic centralized control approaches. A detailed analysis, exploration, and implementation of the high impedance capabilities of distributed controllers for a mobile base with latency-prone centralized processors has not previously been performed.
\begin{figure}
 \centering
   \includegraphics[width = 0.6 \linewidth]{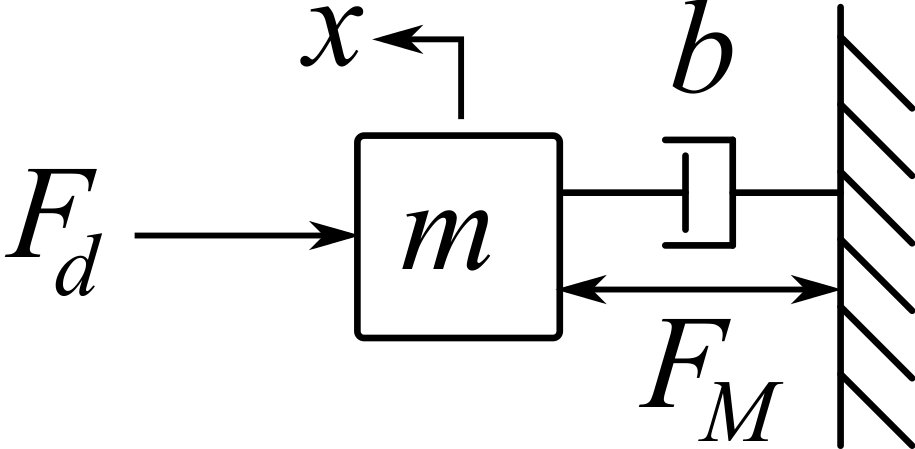}
 \caption{\textbf{Actuator and Control Plant Model.} This diagram represents a generalization of rigid actuators considered in this paper. $F_M$ is the applied motor force, $x$ is the load displacement output, $m$ is effective output inertia, $b$ is the actuator's passive damping, and $F_d$ is an external disturbance force.}
 \label{fig:rigidSchematic}
\end{figure}
\begin{figure*}
 \centering
   \includegraphics[width = 0.8\linewidth]{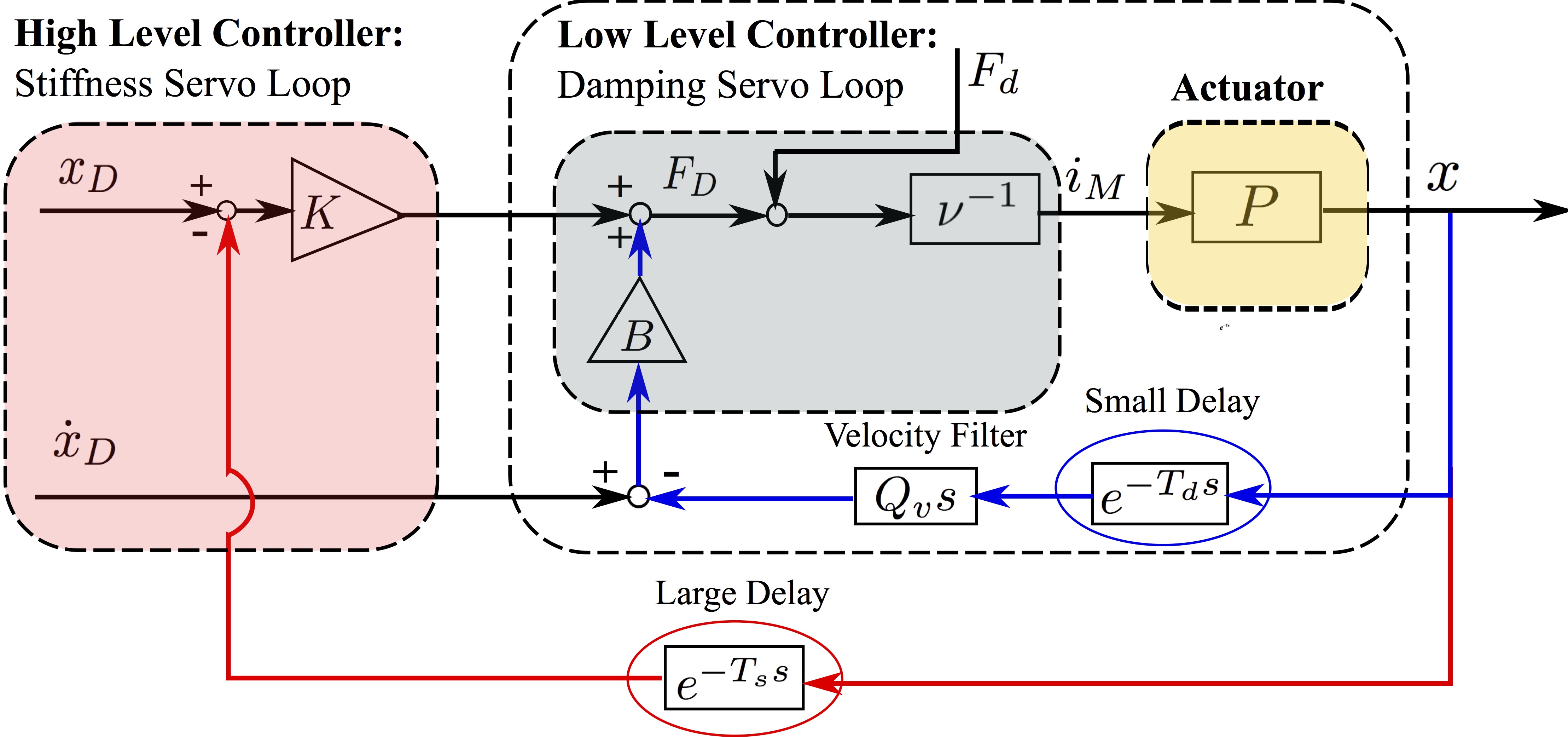}
\caption{\textbf{Single Input / Single Output Controller with Distributed Structure.} A simple proportional-derivative control law is used to control an actuator. $P$ denotes the actuator plant with motor current input, $i_M$, and position output, $x$. $\nu^{-1}$ represents a scaling constant mapping the desired force, $F_D$, to the motor current, $i_M$. $K$ is the stiffness feedback gain while $B$ is the damping feedback gain. The damping feedback loop is labeled as embedded to emphasize that it is meant to be locally implemented to take advantage of high servo rates. On the other hand the stiffness loop is implemented in a high-level computational process close to external sensors and centralized models, for operational space control purposes. Operational space control is normally used in human-centered robotic applications where controllers use task coordinates and global models for their operation. An external disturbance is denoted as $F_d$ inserted between the controller and plant block as suggested by \cite{ogata2010modern}. This simple controller is used to illustrate the discrepancies to latencies between the servo loops. It does not correspond to a practical robot controller as it contains only a single degree of freedom. After studying the physical advantages of these type of structure, we leverage it to a multi-axis robotic base shown in Figure~\ref{fig:Trikey} demonstrating the ability to decouple the damping and stiffness servos to simultaneously achieve system stability and operational space control.}
\label{fig:siso-controller}
\end{figure*}

Robustness and the effects of delay have often been studied in work regarding Proportional-Integral-Derivative (PID) controller tuning. 
A survey of PID controllers including system plants using phase margin techniques with linear approximations is conducted in \cite{lee-pid}. 
The works \cite{astrom-pid, poulin-pid} study auto-tuning and adaption of PID controllers while the work \cite{yaniv-good} furthers these techniques by developing optimal design tools applied to various types of plants including delays. The study in \cite{tipsuwan2004gain} proposed an optimal gain scheduling method for DC motor speed control with a PI controller. In \cite{LeeJC14}, a backstepping controller with time-delay estimation and nonlinear damping is developed for variable PID gain tuning under disturbances. The high volume of studies on PID tuning methods highlight the importance of this topic for robust control under disturbances. However, none of those studies considers the sensitivity discrepancy to latencies between the stiffness and damping servos as separate entities nor do they consider the decoupling of those servos into separate processes for stability purposes as it is done in this paper.

The field of haptics \cite{colgate-max}, networked control \cite{gao2008new, PangZH14} and teleoperation \cite{tipsuwan2004gain} have also thoroughly studied delays and filtering effects. Specifically, haptics is more related to our work because it is a special case of distributed control in which the master and slave devices require separate feedback controllers. Due to the destabilizing effects of time delays, significant effort has been put forth to ensure systems are stable by enforcing passivity criteria \cite{colgate-max}. Other works \cite{lawrence-impedance, colgate-zwidth, hulin2013optimal} further relax this constraint and focus on how delay and filtering affect stability. 
Related work has been performed considering additional real-world effects such as quantization and coulomb friction on system stability \cite{diolaiti-stability}. Once more, theses studies do not analyze nor exploit the large sensitivity discrepancy between stiffness and damping feedback loops nor propose solutions to increase performance based on this discrepancy.


Consequently, our main contribution is to analyze, provide control system solutions, implement and evaluate actuators and mobile robotic systems with latency-prone distributed architectures to significantly enhance their trajectory tracking capabilities under disturbances and system uncertainties. In particular, a new study is performed to reveal that system stability and performance is more sensitive to damping servo latencies than stiffness servo latencies. Then a novel servo breakdown rule is proposed to evaluate the benefits of using a distributed control architecture. As a conclusion, this paper proposes to use stiffness servos for centralized operational space control while realizing embedded-level damping servos as joint space damping processes for stability and tracking accuracy. 


\section{Basic Distributed Control Structure}
\label{sec:control}

This section describes the actuator model used to analyze closed-loop system stability, propose a basic distributed control architecture that delocalizes stiffness and damping servo loops, and analyzes the sensitivity of these control processes to loop delays.

\subsection{Actuator plant model}
Many rigid electrical actuators like the ones used in modern robots can be approximately modeled as a second-order plant with a force acting on an inertia-damper pair (Figure \ref{fig:rigidSchematic}). Considering a current-controlled motor, the control plant from current, $i_M$, to position, $x$, is
\begin{equation}\label{eq:plantTF}
P(s)  = \frac{x(s)}{i_M(s)} = \frac{x(s)}{F_M(s)} \frac{F_M(s)}{i_M(s)} = \frac{\nu}{ms^2 + b s},
\end{equation}
where $F_M$ is the applied motor force, $\nu \triangleq F_M/i_M = \eta N k_{\tau}$, $\eta$ is the drivetrain efficiency, $N$ is the gear speed reduction and $k_{\tau}$ is the motor torque constant.




\subsection{Closed-loop distributed controller model}

Figure \ref{fig:siso-controller} shows our proposed distributed controller built using a proportional-derivative feedback mechanism. It includes velocity feedback filtering ($Q_v s$), stiffness feedback delay ($T_s$), damping feedback delay ($T_d$), with $T_s \neq T_d$, stiffness feedback gain ($K$) and damping feedback gain ($B$). Excluding the unknown load ($F_d$), the desired motor force ($F_D$) in the Laplace domain associated with the proposed distributed controller is
\begin{equation}
F_D(s) = K (x_{D} - e^{-T_s s} x) + B (x_{D}s - e^{-T_d s} Q_v x s),
\end{equation}
where $s$ is the Laplace variable, $x_D$ and $\dot{x}_D$ (i.e., $x_D s$ in the Laplace domain) are the desired output position and velocity, and $e^{-T_s\,s}$ and $e^{-T_d\,s}$ represent Laplace transforms of the time delays in the stiffness and damping feedback loops, respectively. Using the above equation and Equation (\ref{eq:plantTF}), one can derive the closed-loop transfer function from desired to output positions as
\begin{equation}\label{eq:rigidsecondorderTF}
P_{CL}(s) = \frac{x}{x_D} = \frac{B s + K}{m s^2 + (b + e^{-T_ds} B Q_v) s + e^{-T_ss} K},
\end{equation}
where $Q_v$ is chosen to be a first order low pass filter with a cutoff frequency $f_v$
\begin{equation}\label{eq:lpf}
Q_v(s) = \frac{2 \pi f_v}{s + 2 \pi f_v}.
\end{equation}
\begin{figure}
 \centering
   \includegraphics[width=0.85 \linewidth]{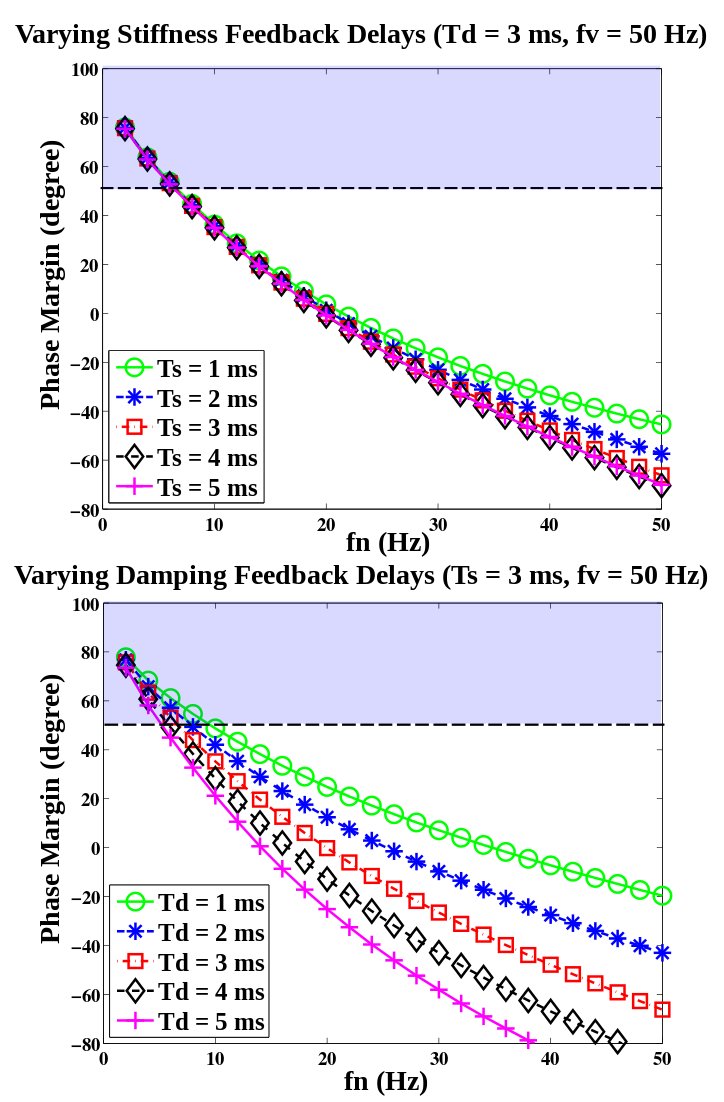}
 \caption{\textbf{Phase Margin Sensitivity to Loop Delays.} This figure shows phase margin simulations of the open loop transfer function shown in (\ref{eq:P_OL}) as a function of the natural frequency defined in (\ref{eq:nat-freq}) and the servo delays shown in Figure~\ref{fig:siso-controller}. A phase margin of $0^\circ$ is considered unstable, however, from simulations of various types of actuators~\cite{paine2014closed}, oscillatory behavior begins below the threshold of $50^\circ$, displayed above as a horizontal line. Delays ranging between $1$ ms and $5$ ms are simulated for both the stiffness and damping servos. The simulations are performed based on identical actuator parameters than those used in the experimental section, Section \ref{sec:experiments}, i.e. passive output inertia $m = 256$ kg and passive damping $b = 1250$ Ns/m. Equations~(\ref{eq:secondcriticaldamp}) and (\ref{eq:nat-freq}) can subsequently be used to derived the stiffness and damping feedback gains. The key observation here is that variations of the phase margin curves above are much more pronounced from damping servo delays than from stiffness servo delays.}
 \label{fig:Sensitivity1}
\end{figure}
To derive the open-loop transfer function~\cite{ogata2010modern} of the distributed controller, one can re-write Equation ({\ref{eq:rigidsecondorderTF}}) as
\begin{equation}\label{eq:closed-loop}
P_{CL}(s) = \frac{\frac{B s + K}{m s^2 + b s}}{1+ P_{OL}(s)},
\end{equation}
where $P_{OL}(s) \triangleq P(s) H(s)$ is the open-loop transfer function,
\begin{equation}\label{eq:P_OL}
P_{OL}(s) =\frac{e^{-T_d s} B Q_v s + e^{-T_s s}K}{m s^2 + b s},
\end{equation}
$P(s)$ is the actuator's plant, and $H(s)$ is the so-called feedback transfer function.

The presence of delays and filtering causes the above closed loop plant to behave as a high order dynamic system for which typical gain selection methods do not apply. However, to make the problem tractable, one can define a dependency between the stiffness and damping gains using an idealized second order characteristic polynomial \cite{ogata2010modern}
\begin{equation}\label{eq1}
s^2 + 2\zeta\omega_n s + \omega_n^2,
\end{equation} 
where $\omega_n$ is the so-called natural frequency and $\zeta$ is the so-called damping factor. 
In such case, the idealized characteristic polynomial (i.e. ignoring delays, $T_s = T_d = 0$, and filtering, $Q_v = 1$) associated with our closed loop plant of Equation~(\ref{eq:rigidsecondorderTF}) would be
\begin{equation}\label{eq2}
s^2 + (B+b)/m \cdot s + K/m.
\end{equation}
Choosing the second order critically damped rule, $\zeta = 1$ and comparing Equations~(\ref{eq1}) and (\ref{eq2}) one can get the gain dependency
\begin{equation}\label{eq:secondcriticaldamp}
B = 2 \, \sqrt{mK} - b,
\end{equation}
and the natural frequency,
\begin{equation}\label{eq:nat-freq}
f_n \triangleq \frac{\omega_n}{2 \pi} = \frac{1}{2 \pi} \sqrt{\frac{K}{m}}.
\end{equation}
The second order dependency of Equation~(\ref{eq:secondcriticaldamp}) will be used for the rest of this paper for deriving new gain selection methods through the thorough analysis of the oscillatory behavior of the closed loop plant of Equation~(\ref{eq:rigidsecondorderTF}). In particular our study will use the phase margin criterion and other visualizations tools to study how the complete system reacts to feedback delays and signal filtering. Phase margin is the additional phase value above $-180^\circ$ when the magnitude plot crosses the 0 dB line (i.e., the gain crossover frequency). It is common to quantify system stability by its phase margin.

\begin{figure*}
 \centering
   \includegraphics[width=0.9\linewidth]{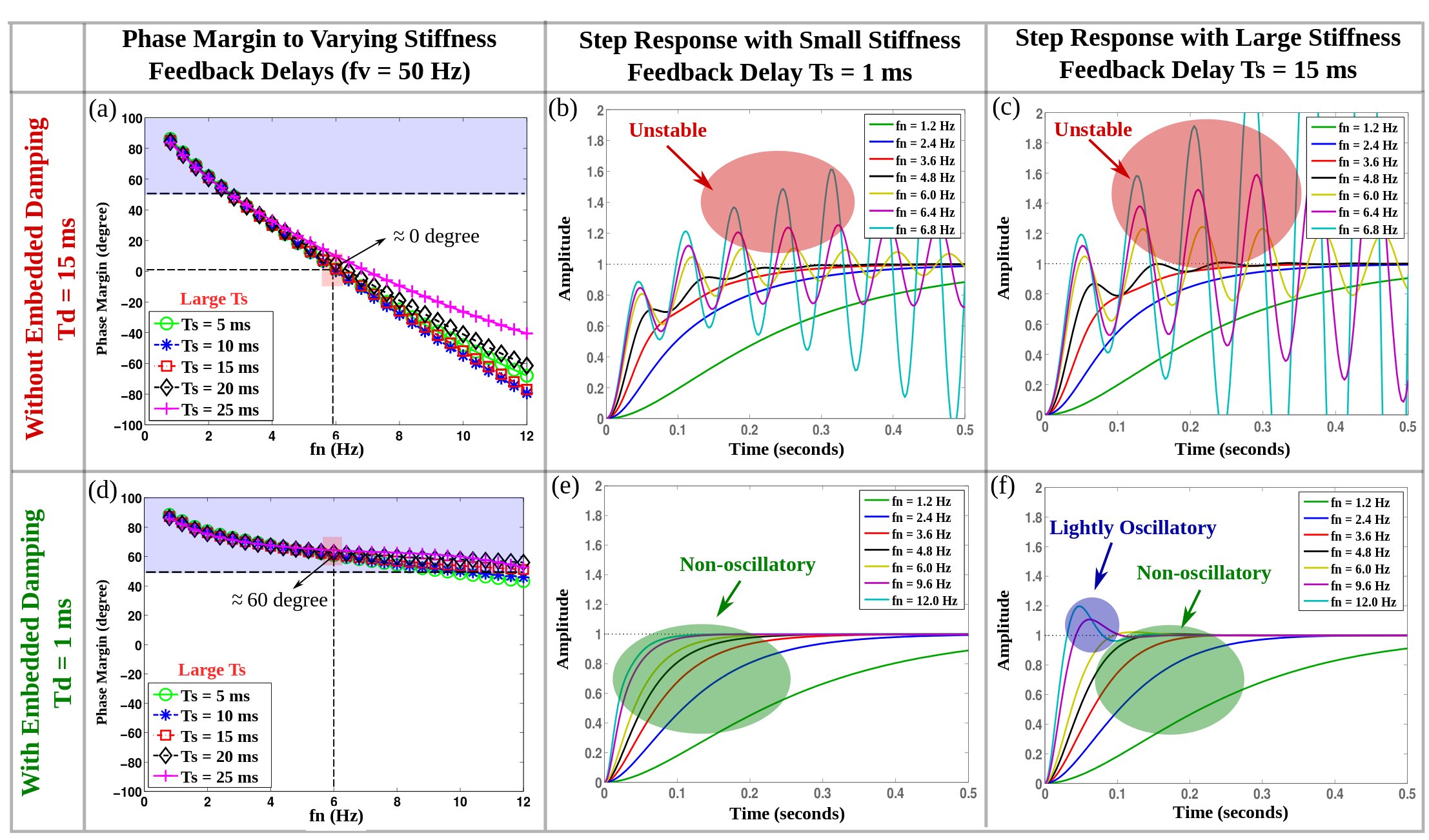}
 \caption{\textbf{Comparison of Step Response with Slow and Fast Damping Servos.} The figures above compare the effects of implementing damping feedback on slow or fast servo processes. The simulations are performed on the same actuator used in the experimental section, Section \ref{sec:experiments}. The top row depicts damping feedback implemented with delays of $15$ ms while the bottom row depicts a faster damping servo with delay of only $1$ ms. For both rows, various stiffness delays are identically used ranging from $5$ ms to $25$ ms. Subfigures (a) and (d) show simulations of the phase margin as a function of the natural frequency, which in turn is a function of the feedback stiffness gain. The first point to notice here is that the phase margin values for subfigure (a) are significantly lower than for (d) due to the larger damping delay. Secondly, both (a) and (d) show small variations between the curves, corroborating the small sensitivity to stiffness delays that will be studied in Section~\ref{sec:causes}. Corresponding step responses are shown along for various natural frequencies. 
The key observation here is that embedded damping greatly improves the stability of the controller despite large stiffness delays.}
 \label{fig:StepResponseED}
\end{figure*}


\subsection{Phase margin sensitivity comparison}
\label{sec:embeddamp}

This subsection focuses on utilizing frequency domain control method to analyze the phase margin sensitivity to time delays on the distributed control architecture shown on Figure~\ref{fig:siso-controller}. Different delay range scales are considered: (1) a small range scale ($1-5$ ms) to show detailed variations, and (2) a larger range scale ($5-25$ ms) to cover practical control ranges. These scales roughly correspond to embedded and centralized computational and communication processes found in highly articulated robots such as \cite{paine2014actuator}. Phase margin plots, are subsequently obtained for the controller of Equation~(\ref{eq:rigidsecondorderTF}) and shown on Figures~\ref{fig:Sensitivity1} and \ref{fig:StepResponseED} as a function of the natural frequency given in Equation~(\ref{eq:nat-freq}) and using the gain relationship of Equation~(\ref{eq:secondcriticaldamp}). 


From Figure~\ref{fig:Sensitivity1}, it is noticeable that reducing either stiffness or damping feedback delays will increase the stability of the controller. But more importantly, it is clearly visible that phase margin behavior is much more sensitive to damping servo delays ($T_d$) than to stiffness servo delays ($T_s$). Not only there is a disparity on the behavior with respect to the delays, but phase margin is fairly insensitive to stiffness servo delays in the observed time scales. Such disparity and behavior is the central observation that motivates this paper and the proposed distributed control architecture. Figure~\ref{fig:StepResponseED} simulates step position response of the controller for a range of relatively large stiffness delays and for two choices of damping delays, a short and a long one. It becomes clear that reducing damping delay significantly boosts stability even in the presence of fairly large stiffness delays. These results emphasize the significance of implementing damping terms at the fastest possible level (e.g. at the embedded level) while proportional (i.e. stiffness) servos can run in latency prone centralized processes.

\section{Basis for Sensitivity Discrepancy}
\label{sec:causes}

In the previous section it was observed different behavior of the controller's phase margin depending on the nature of delay. Damping delay seems to affect much more the system's phase margin than stiffness delay. This section will analyze this physical phenomenon in much more detail and reveal the conditions under which this disparity occurs.

\subsection{Equations expressing phase margin sensitivity to delays}

Detailed mathematical analysis is developed to find further physical structure for the causes of stability discrepancies between damping and stiffness delays. Let us re-visit the open loop transfer function of Equation~(\ref{eq:P_OL}). The resulting open loop transfer function, including the low pass velocity filter of Equation~(\ref{eq:lpf}), in the frequency domain ($s = j\omega$) is
\begin{equation}\label{eq:P_OLomega}
P_{OL}(\omega) = \frac{j A_1(\omega) + A_2(\omega)}{j\omega(jm\omega + b)(j\tau_v\omega + 1)},
\end{equation}
with
\begin{align}\nonumber
A_1(\omega) &\triangleq B \, \omega \,{\rm cos}(T_d\;\omega) - K\,{\rm sin}(T_s\;\omega) + K \tau_v \omega \, {\rm cos}(T_s \omega),\\\label{eq:As}
A_2(\omega) &\triangleq B \, \omega \,{\rm sin}(T_d\;\omega) + K\,{\rm cos}(T_s\;\omega) + K \tau_v \omega \, {\rm sin}(T_s \omega).
\end{align}
Note that Euler's Formula ($e^{-jx} = {\rm cos}x - j{\rm sin}x$) has been used to obtain the above results.

The phase margin, $PM \triangleq 180^\circ + \angle P_{OL}(\omega_g)$, of the plant (\ref{eq:P_OLomega}), where $\angle .$ is the angle of the argument, is 
\begin{equation}
PM = \rm{atan}\left[\frac{A_{1g}}{A_{2g}}\right] + 90^\circ - \rm{atan}\left[\frac{m \, \omega_g}{b}\right] - \rm{atan}\left[\tau_v \omega_g\right],
\end{equation}
with $\omega_g$ being the gain crossover frequency~\cite{ogata-modern} and $A_{ig} \triangleq A_i(\omega_g), i = \{1,2\}$.

Following the derivations of Appendix~\ref{sec:appendix1}, one can obtain the sensitivity equations below expressing variations of the phase margin with respect to stiffness and damping delays,
\begin{gather}\label{eq:sens1}
\frac{\partial PM}{\partial T_s} = \frac{\left[ -K^2 (\tau_v^2 \omega_g^2 + 1) + K \, B \, \omega_g \; M \right] \, \omega_g}{A_{1g}^{\;2} + A_{2g}^{\;2}},\\[3mm]\label{eq:sens2}
\frac{\partial PM}{\partial T_d} = \frac{\left[ -B^2 \omega_g^2 + K \, B \, \omega_g \; M \right]\, \omega_g}{A_{1g}^{\;2} + A_{2g}^{\;2}}.
\end{gather}
where $M = \sqrt{(\tau_v \omega_g)^2 + 1} \cdot {\rm sin}\Big((T_s-T_d) \;\omega_g + \phi \Big)$ is also derived in that Appendix~\ref{sec:appendix1}.

\subsection{Gain crossover sensitivity condition}

From the control analysis of the distributed plant performed in previous sections, increasing damping delays decreases the phase margin. This observation means that the sensitivity of the phase margin to damping delays must be negative, i.e.
\begin{equation}\label{eq:prop1}
\frac{\partial PM}{\partial T_d} < 0.
\end{equation}
Also from those analysis, it is observed that the phase margin is more sensitive to damping than to stiffness delays. This observation can be formulated as
\begin{equation}\label{eq:prop2}
\frac{\partial PM}{\partial T_d} < \frac{\partial PM}{\partial T_s}.
\end{equation}

Let us re-organize the numerator of Equation~(\ref{eq:sens2}) to be written in the alternate form
\begin{equation}
\frac{\partial PM}{\partial T_d} = \frac{\left[ -B \omega_g + K M \right] B \, \omega_g^2}{A_{1g}^{\;2} + A_{2g}^{\;2}}.
\end{equation}
An upper bound of the above equation occurs when the maximal condition ${\rm sin}\Big((T_s-T_d) \;\omega_g + \phi \Big) = 1$ is met, i.e.
\begin{equation}\label{eq:bound}
\frac{\partial PM}{\partial T_d} \leq \frac{\left[ -B \omega_g + K \sqrt{(\tau_v \omega_g)^2 + 1} \right] B \, \omega_g^2}{A_{1g}^{\;2} + A_{2g}^{\;2}}.
\end{equation}
Based on the above inequality, (\ref{eq:prop1}) is met if the following gain crossover sensitivity condition is met,
\begin{equation}\label{eq:prop3}
\omega_g > \frac{K}{\sqrt{B^2 - K^2 \tau_v^2}}.
\end{equation}
The above equation is only a sufficient condition for fulfilling Condition~({\ref{eq:prop1}}). Obtaining a closed form solution for that condition would be very complex due to the presence of trigonometric terms. Therefore, the remainder of this section is to study under what circumstances Condition~({\ref{eq:prop3}}) holds.

At the same time, Inequality~(\ref{eq:prop2}) can be re-written in the form
\begin{equation}
\frac{\partial PM}{\partial T_d} - \frac{\partial PM}{\partial T_s} = 
\frac{\left[ -B^2 \omega_g^2 + K^2 (\tau_v^2 \omega_g^2 + 1) \right] \omega_g}{A_{1g}^{\;2} + A_{2g}^{\;2}}
< 0,
\end{equation}
where it has been subtracted the right hand sides of Equations~(\ref{eq:sens1}) and (\ref{eq:sens2}) for the derivation. Notice that in that subtraction the sine functions cancel out. Coincidentally, the above inequality is also met if the gain crossover sensitivity condition (\ref{eq:prop3}) is fulfilled. In other words, that condition is sufficient to meet both Inequalities~(\ref{eq:prop1}) and (\ref{eq:prop2}). 

\subsection{Servo breakdown gain rule}

To validate the gain crossover condition~(\ref{eq:prop3}), our study solves for the gain crossover frequency, which consists of the frequency at which the magnitude of the open loop transfer function is equal to unity, i.e.
\begin{equation}\label{eq:unit-gain}
|P_{OL}(\omega_g)| = 1.
\end{equation}
Using the plant~(\ref{eq:P_OLomega}), it can be shown that the above equation results in the equality (see the Appendix~\ref{sec:appendix2} for the derivations),
\begin{align}\label{eq:fourthorderequation}\nonumber
&(B\omega_g )^2 + K^2 (\tau_v^2 \omega_g^2 + 1) - 2 KB\omega_g M \\
&= \omega_g^2 \Big((\omega_g m)^2 + b^2\Big)\Big(\tau_v^2 \omega_g^2 + 1\Big).
\end{align}
The above equation is intractable in terms of deriving a closed loop expression of the gain crossover frequency. To tackle a solution our study introduces transformations of the parameters and numerically derives parameter ranges for which Condition~({\ref{eq:prop3}}) holds. Let us start by creating a new variable that allows to write ({\ref{eq:prop3}}) as an equality,
\begin{equation}\label{eq:omega_g}
\delta \in [-1,\infty) \quad s.t. \quad \omega_g = (1+\delta) \, \frac{K}{\sqrt{B^2 - K^2 \tau_v^2}}.
\end{equation}

Thus, demonstrating the gain crossover sensitivity condition~{(\ref{eq:prop3})} is equivalent to demonstrating that $\delta>0$. Rewriting Equation~{(\ref{eq:secondcriticaldamp})} as $K = (B+b)^2/4m$ and substituting $K$ in the above equation, {(\ref{eq:omega_g})} can be further expressed as
\begin{equation}\label{eq:omega_g2}
\omega_g = (1+\delta) \, \frac{(B+b)^2}{\sqrt{16 B^2 m^2 - (B + b)^4 \tau_v^2}}.
\end{equation}
Dividing Equation~{(\ref{eq:fourthorderequation})} by a new term $K^2\,U\,V$, with $U \triangleq \tau_v^2 \omega_g^2 +1$, and $V 
\triangleq B^2 \, \omega_g^2 / K^2$, while substituting $\omega_g$ on the right hand side of Equation~(\ref{eq:fourthorderequation}) by Equation~(\ref{eq:omega_g2}), and using $M$ as shown in Equation~(\ref{eq:MM}), Equation {(\ref{eq:fourthorderequation})} becomes
\begin{multline}\label{eq:deltas}
\frac{1}{U} + \frac{1}{V} - \frac{2 {\rm sin}\Big((T_s - T_d)\,\omega_g + \phi\Big)}{\sqrt{U\cdot V}}\\
= \frac{\big(1+\delta\big)^2\,\big(B+b\big)^4}{16\,B^4 - B^2 (B + b)^4 \tau_v^2/m^2} + \left(\frac{b}{B}\right)^2.
\end{multline}
Using Equation~(\ref{eq:omega_g}) it can be further demonstrated that $V = (\tau_v \omega_g)^2 + (1 + \delta)^2$. Thus $U$ and $V$ are only expressed in terms of $(\tau_v \omega_g)^2$. To further facilitate the analysis, let us introduce three more variables
\begin{gather}\label{eq:alpha}
\alpha \triangleq {\rm sin}\bigg((T_s - T_d)\,\omega_g + \phi \bigg)\,\in [-1,1],\\\label{eq:beta}
\beta \in (0,\infty) \quad s.t. \quad B = \beta\,m,\\\label{eq:gamma}
\gamma \in (0,\infty) \quad s.t. \quad B = \gamma\,b.
\end{gather}
Notice that $\alpha$ can be interpreted as an uncertainty, $\beta$ is the ratio between damping gain and motor drive inertia and $\gamma$ is the ratio between damping gain and motor drive friction. Using these variables, (\ref{eq:deltas}) simplifies to
\begin{equation}\label{eq:simplified}
\frac{U+V - 2 \alpha \sqrt{U \cdot V}}{U \cdot V}
= \frac{\big(1+\delta\big)^2\,\big(1+\gamma)^4}{16\,\gamma^4 - (1 + \gamma)^4 \beta^2 \tau_v^2} + \frac{1}{\gamma^2}.
\end{equation}
Using Equations~({\ref{eq:omega_g2}}),~({\ref{eq:beta}}) and~({\ref{eq:gamma}}), the term $(\tau_v \omega_g)^2$ appearing in the variables $U$ and $V$ on Equation~({\ref{eq:simplified}}) can be expressed as
\begin{align}\label{eq:filter_term-2}
(\tau_v \omega_g)^2& = \beta^2 \tau_v^2 \frac{\big(1+\delta\big)^2\,\big(1+\gamma)^4}{16\,\gamma^4 - (1+\gamma)^4 \beta^2 \tau_v^2}
\end{align}
Thus, Equation~({\ref{eq:simplified}}) does not contain direct dependencies with $\omega_g$ and therefore can be represented as the nonlinear function
\begin{equation}
f(\alpha, \beta, \gamma, \delta, \tau_v) = 0
\end{equation}

\begin{figure}
 \centering
   \includegraphics[width=\linewidth]{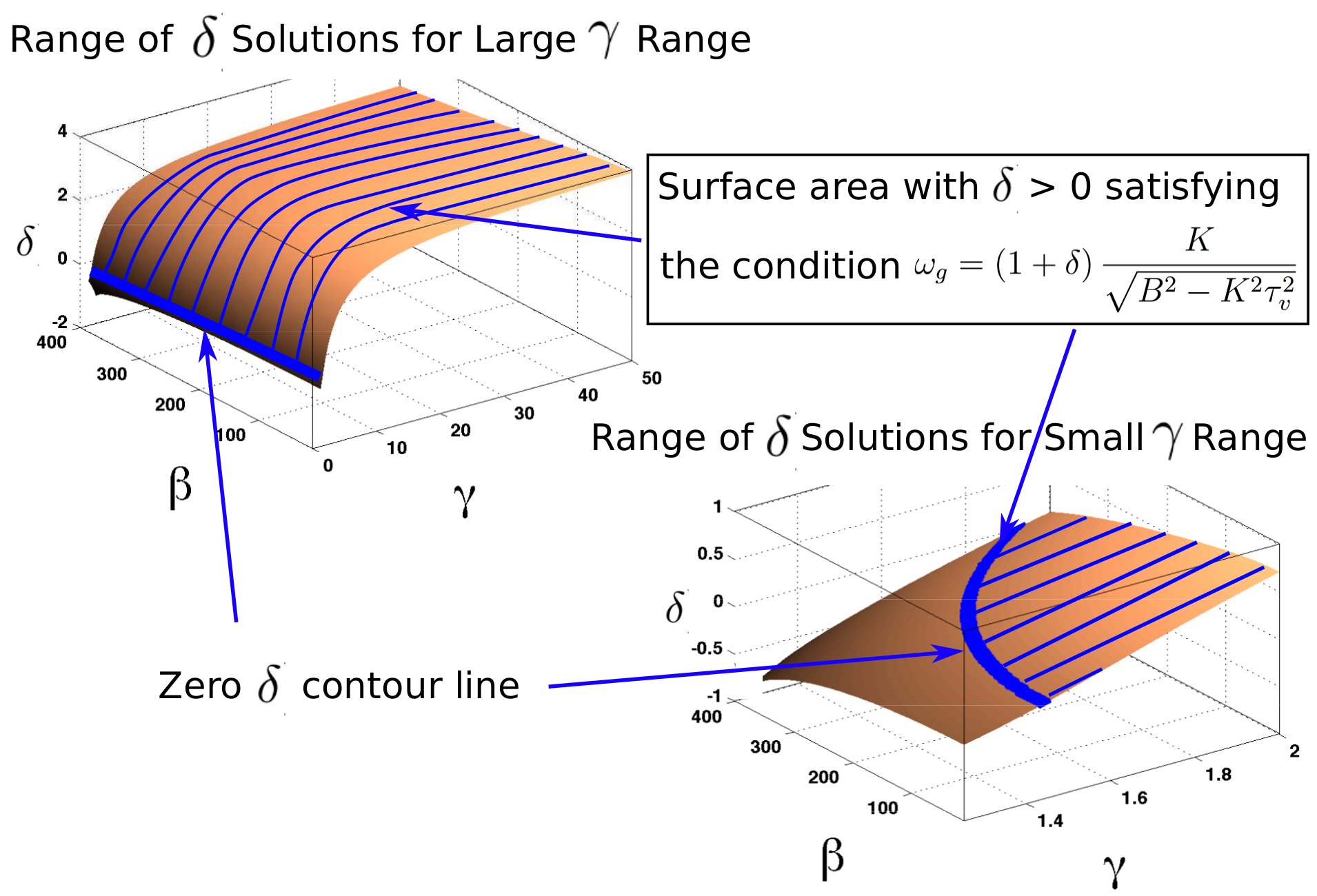}
 \caption{\textbf{Controller Values Meeting the Gain Crossover Sensitivity Condition.} The surfaces above show the range of feedback parameters that meet the gain crossover sensitivity condition of Equation~(\ref{eq:prop3}). $\delta > 0$ represents the excess gain ratio by which the condition is met. $\gamma > 0$ represents the ratio between damping feedback gain and passive damping. $\beta \in [10, 400]$ is chosen to cover a wide range of actuator parameters. The surfaces above demonstrate that a wide range of practical gains, $\gamma$, meet the aforementioned gain crossover sensitivity condition. The values of the above surfaces are solved by numerically identifying the smallest real root of Equation~(\ref{eq:simplified}). In the bottom right surface, it can be seen that $\delta>0$ for $\gamma > 2$, meaning that the gain crossover sensitivity condition is met if the ratio between damping feedback gain and passive damping is larger than two.}
 \label{fig:deltasolutions}
\end{figure}

Let us demonstrate under which conditions $\delta > 0$, which will imply that Equation~({\ref{eq:prop3}}) holds. 
In our lab, velocity filters with $\tau_v = 0.0032 \,s$ are commonly used for achieving high performance control \cite{paine2014closed}, and therefore Equation~{(\ref{eq:simplified})} will be solved for only that filter. Notice that it is not difficult to try new values of $\tau_v$ when needed. Additionally, when sampling Equation~{(\ref{eq:simplified})} for the values of $\alpha$ shown in Equation~{(\ref{eq:alpha})} we observed that not only $\delta$ is fairly invariant to $\alpha$ but the lowest value of $\delta$ occurs for $\alpha =1$. These behaviors are omitted here for space purposes. Therefore, as a particular solution, Equation~{(\ref{eq:simplified})} is solved for the values
\begin{equation}
f(\alpha =1 , \beta, \gamma, \delta, \tau_v =0.0032) = 0.
\end{equation}
The above function is solved numerically and the solution surface is plotted in Figure~{\ref{fig:deltasolutions}}. As it can be seen, $\delta>0$ for $\gamma > 2$, allowing us to state that using a distributed PD feedback control law like the one in Figure~{\ref{fig:siso-controller}} with the particular choice of the filter $\tau_v = 0.0032\,s$ and with damping gains greater than
\begin{equation}\label{eq:dampingratio}
B\,>\,2\,b,
\end{equation}
causes the phase margin to be more sensitive to damping delays than to stiffness delays. The threshold above can therefore be interpreted as a breakdown gain rule which is sufficient to meet the gain crossover sensitivity condition~(\ref{eq:prop3}), and from which the aforementioned phase margin sensitivity discrepancy follows.

This threshold hints towards a general rule for breaking controllers down into distributed servos, as was illustrated in Figure~\ref{fig:siso-controller}. Namely, if the maximum allowable feedback damping gain for a given servo rate is significantly larger than twice the passive actuator damping, then the controller's stiffness servo can be decoupled from the damping servo to a slower computational process without hurting the controller's stability.



\subsection{An example: analyzing real-world actuators by the servo breakdown rule}

As a means of demonstrating the utility of the breakdown gain rule of Equation~(\ref{eq:dampingratio}), here we analyze several real-world actuation systems. Our goal is to determine whether the properties of each system make them good candidates for distributed control schemes with decoupled stiffness and damping feedback loops.

\begin{figure}
 \centering
   \includegraphics[width=0.9\linewidth]{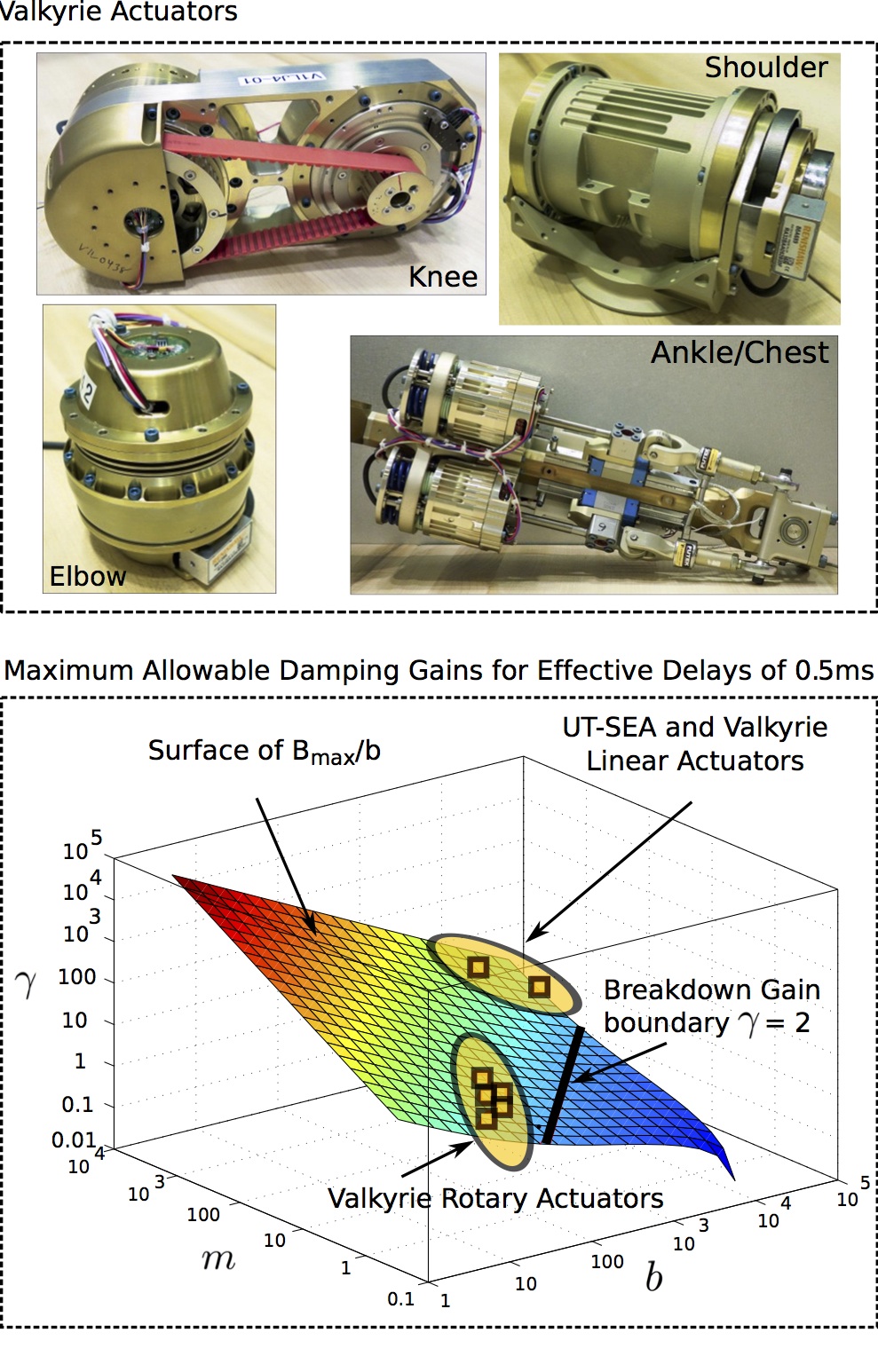}
 \caption{\textbf{Various Actuators Meeting Servo Breakdown Rule.} The top image shows various actuators from NASA that our group helped to build. The bottom image shows the surface of maximum allowable damping gains (computed according to the gain rule described in~\cite{paine2014closed}) as a function of actuator parameters. $\gamma$ is the feedback damping gain ratio described in Equation~(\ref{eq:gamma}), and $m$ and $b$ are the output inertia and passive damping of the various actuators. As it is shown, the maximum allowable gains are above the servo breakdown boundary. This entails that stiffness servos are fairly insensitive to delays and therefore can be decoupled from the damping servos when needed.}
 \label{fig:DifferentParameterRange}
\end{figure}

Table~\ref{table:Actuator} shows actuator parameters for the Valkyrie humanoid and the UT-SEA actuator~\cite{painedesign2014}, as well as the maximum feedback damping gains that are implemented in those actuators to achieve maximum impedance control. Our lab has been involved in developing these two sets of actuators.
In all instances, the embedded servos had effective delays of $0.5$ ms. In order to compute the maximum feedback damping gains as a function of the previous servo rate, our recent work~\cite{paine2014closed} is used. 
In that work, a new rule is provided to compute maximum feedback gains for a phase margin of $50^\circ$ given the actuator parameters and the servo rate. 
Figure~\ref{fig:DifferentParameterRange} shows, (1) pictures of various Valkyrie actuators, (2) a surface depicting the maximum allowable damping gains as a function of actuator parameters, and (3) Valkyrie's actuators mapped into the surface. The surface is computed for effective delays of $0.5$ ms. Within the surface, it shows the line corresponding to the breakdown gain rule of Equation~(\ref{eq:dampingratio}). 
As can be seen, all actuators implement feedback damping gains that were above the breakdown gain boundary. It follows that those gains would be highly sensitive to damping servo delays. 

To maintain these maximum actuator gains, servo latency for the damping process must not be increased. However, according to the servo breakdown rule, the stiffness servo processes shall be fairly insensitive to delays and therefore could be decoupled and implemented in a slower centralized process. Such decoupling is advantageous in multi-axis robots where centralized processes contain sensor and model information needed for operational space control to coordinate the robot's movement. In Subsection~\ref{subsec:doscmb} we discuss such an application for an omnidirectional mobile robotic base.
\begin{table}
\caption{UT-SEA/Valkyrie Actuator Parameters} 
\centering
\begin{tabular}{|c||c|c|c|c|}
\hline
\hline
Actuator& output inertia & passive damping & damping gain & ratio\\ 
Type &$m$ & $b$ & B & $\gamma$ \\ \hline \hline
UT-SEA & 360 kg & 2200 N$\cdot$s/m & 50434 N$\cdot$s/m & 22.92\\  \hline 
Valkyrie 1 & 270 kg & 10000 N$\cdot$s/m & 46632 N$\cdot$s/m & 4.66 \\ \hline
Valkyrie 2 & 0.4 kg$\cdot$m$^2$ & 15 Nm$\cdot$s/rad  & 68 Nm$\cdot$s/rad & 4.55 \\ \hline
Valkyrie 3 & 1.2 kg$\cdot$m$^2$ & 35 Nm$\cdot$s/rad& 196  Nm$\cdot$s/rad & 5.60 \\  \hline 
Valkyrie 4 & 0.8 kg$\cdot$m$^2$ & 40 Nm$\cdot$s/rad& 145  Nm$\cdot$s/rad & 3.61\\  \hline 
Valkyrie 5 & 2.3 kg$\cdot$m$^2$ & 50 Nm$\cdot$s/rad& 360  Nm$\cdot$s/rad & 7.20\\  \hline 
Valkyrie 6 & 1.5 kg$\cdot$m$^2$ & 60 Nm$\cdot$s/rad& 259  Nm$\cdot$s/rad & 4.32\\  \hline 
\end{tabular}
\label{table:Actuator}
\end{table}

\section{Experimental Evaluation}
\label{sec:experiments}

\begin{figure*}
 \centering
   \includegraphics[width=\linewidth]{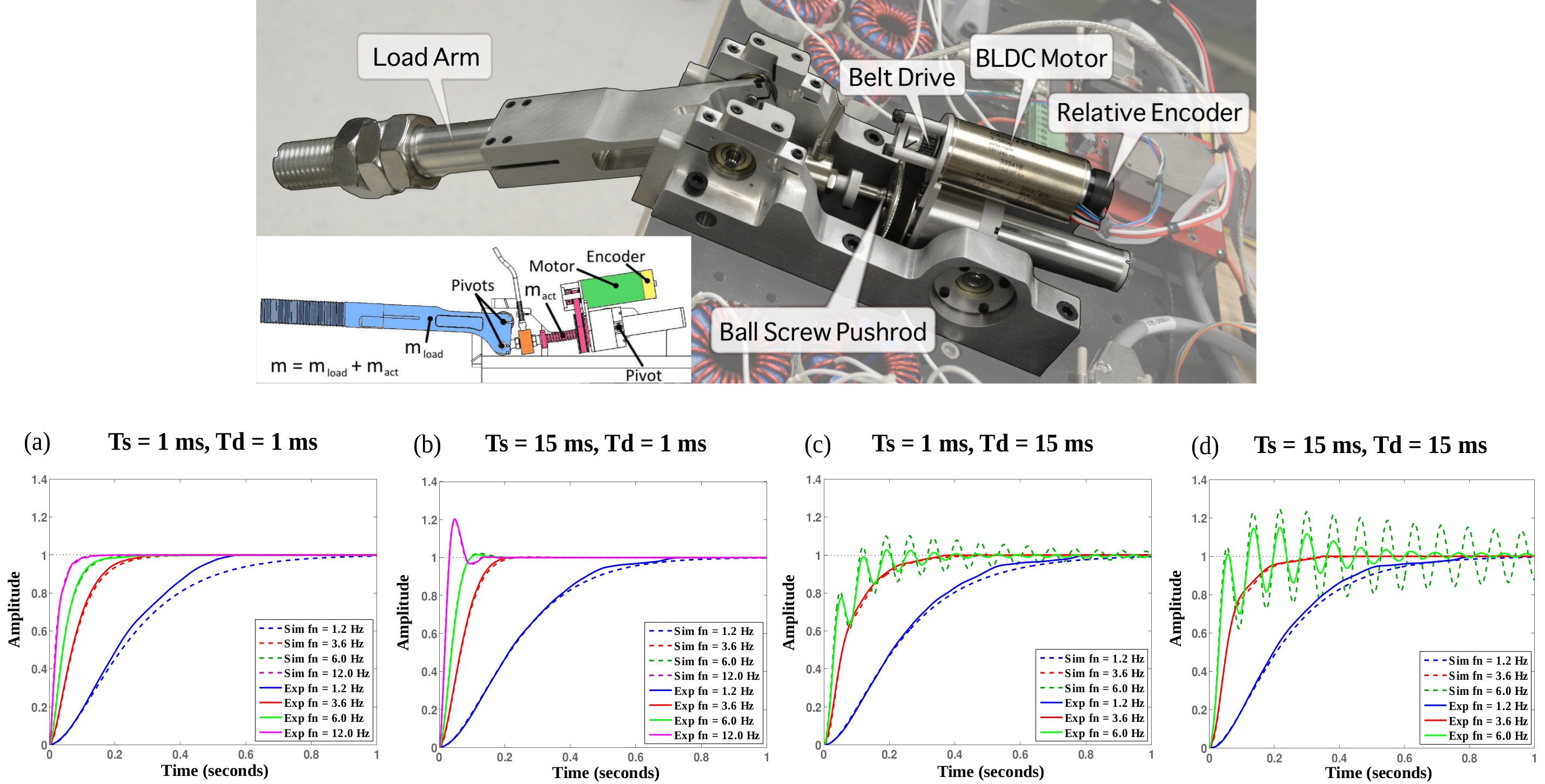}
 \caption{\textbf{Step Response Experiment with Distributed Controller.} Subfigures (a) through (d) show various implementations on the UT linear rigid actuator corresponding to the simulations depicted on Figure \ref{fig:StepResponseED}. Overlapped with the data plots, simulated replicas of the experiments are also shown to validate our models. The experiments not only confirm the higher sensitivity of the actuator to damping than to stiffness delays but also indicate a good correlation between the real actuator and the simulations.}
 \label{fig:StepExpSimu_withoutweight}
\end{figure*}

The proposed controller of Figure~\ref{fig:siso-controller} is implemented in our UT linear rigid actuator shown in Figure~\ref{fig:StepExpSimu_withoutweight}. This linear pushrod actuator has an effective output inertia of $m = 256$ kg and an approximate passive damping of $b = 1250$ Ns/m. The sampling period is $0.5$ ms, i.e., with a $2$ kHz servo rate. At the same time, the controller is simulated by using the closed loop plant given in Equation~(\ref{eq:closed-loop}). Identical parameters to the real actuator are used for the simulation, thus allowing us to compare both side by side.

\subsection{Step response implementation}
First, a test is performed on the actuator evaluating the response to a step input on its position. The results are shown in the bottom part of Figure~\ref{fig:StepExpSimu_withoutweight} which shows and compares the performance of the real actuator versus the simulated closed loop controller. 
Various tests are performed for the same reference input with varying time delays. In particular large and small delays are used for either or both the stiffness and damping loops. The four combinations of results are shown in the figure with delay values of $1$ ms or $15$ ms.

The first thing to notice is that there is a good correlation between the real and the simulated results both for smooth and oscillatory behaviors. Small discrepancies are attributed to unmodelled static friction and the effect of unmodelled dynamics.
More importantly, the experiment confirms the anticipated discrepancy in delay sensitivity between the stiffness and damping loops. Large servo delays on the stiffness servo, corresponding to subfigures (a) and (b) have small effects on the step response. On the other hand, large servo delays on the damping servo, corresponding to subfigures (c) and (d), strongly affect the stability of the controller. In fact, for (c) and (d) the results corresponding to $f_n = 12$ Hz are omitted due to the actuator quickly becoming out of control. In contrast, the experiment in (b) can tolerate such high gains despite the large stiffness delay.


\vspace{-0.05in}
\subsection{Trajectory tracking with an uncertain load}

\begin{figure*}
 \centering
    \includegraphics[width=\linewidth]{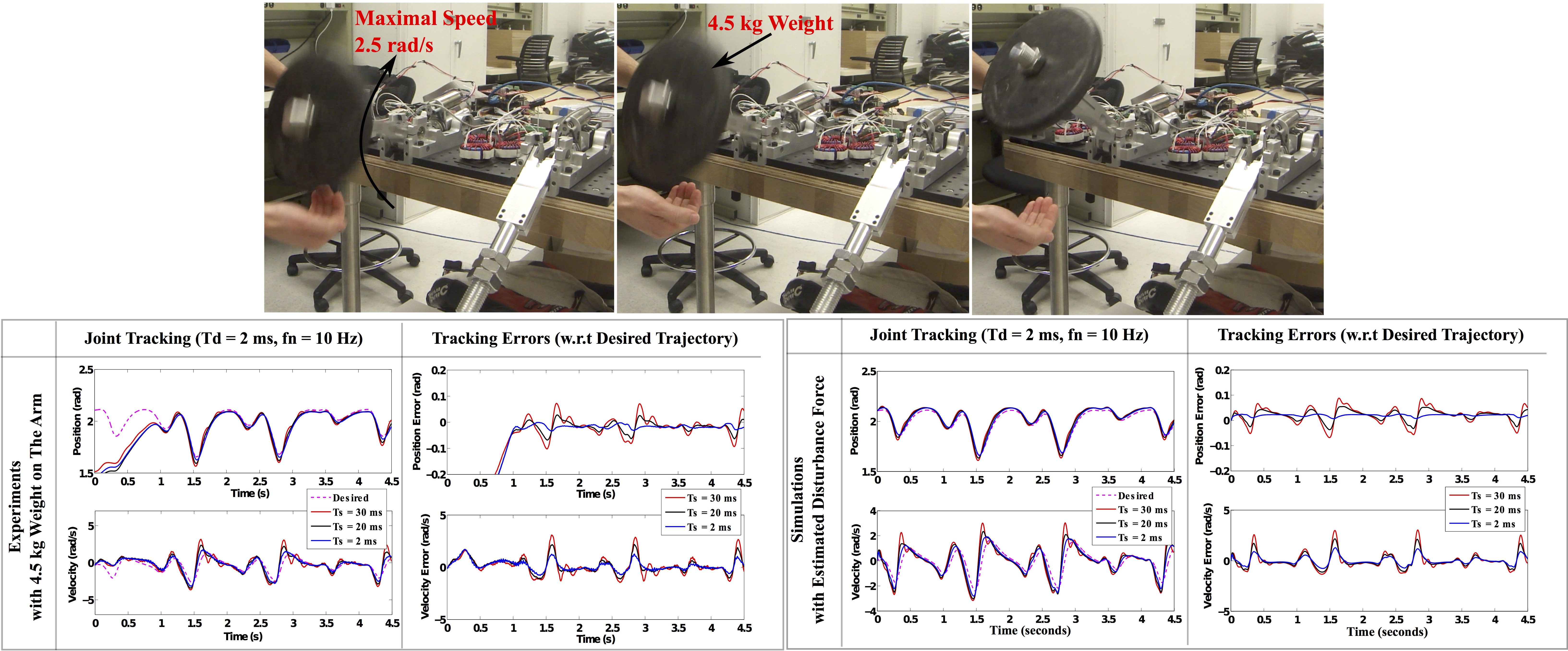}
 \caption{\textbf{Trajectory Tracking Experiment under Load Disturbances with Distributed Controller.} The figures on the top row show snapshots of the testbed with an unmodelled load of $4.5$ kg. The plots show trajectory tracking performance for a small damping delay, $T_d = 2$ ms and various stiffness delays ranging from $T_s = 2$ ms to $T_s = 30$ ms. Trajectory tracking errors on the bottom left remain relatively small despite the large stiffness delays, confirming the advantages of implementing damping feedback in a fast computational processes. A simulated experiment is also shown on the bottom right confirming good correlation between the real and simulated performance.}
 \label{fig:SlowExpandSimu}
\end{figure*}

Performance limits are explored at their fullest in the test shown in Figure~\ref{fig:SlowExpandSimu}. Here, an unmodelled weight of $4.5$kg is attached to the load arm which is also connected to the pushrod actuator through a pivot. The weight is unmodelled and therefore constitutes a disturbance. 
By estimation, the total disturbance torque that the controller has to deal with is $F_d = 16.84$ Nm. A trajectory with output angle variations within [$86^\circ, 126^\circ$]
is designed to test the controller's performance under the load disturbance. This trajectory is inspired by that of a biped locomotion knee joint motion \cite{zhao2012three} during fast walking, with angular velocities varying between $\pm2.5$ rad/s.

This experiment tests the tracking performance under the load disturbance on both the real actuator and also on a numerical simulation of the controller model depicted in Figure~\ref{fig:siso-controller}. Disturbance forces for the numerical simulation are applied based on the position of the arm and considering only gravitational effects. Judging from the visualization of the errors in that figure and the root mean square of the errors depicted in Table~\ref{table:RMSError}, there is a good correlation between the real experiment and the simulation values for both the joint positions and angular velocities.

Once more, the test confirms the predicted robustness to stiffness delays that was studied in previous sections. Increasing $10$ times $T_s$ from $2$ ms to $20$ ms only increases the RMS joint position and angular velocity errors by less than two fold as shown in Table~\ref{table:RMSError}.
By using high gains, the controller ensures that joint tracking is accurate despite the large load disturbances. As shown in the previous table, the maximum root mean square (RMS) error for the tracked joint position is $0.0182$ rad $\approx 1^\circ$ for stiffness delays of $T_s = 2$ ms and $0.036$ rad $\approx 2^\circ$ for delays $T_s=30$ ms.

\begin{table}
\caption{Root Mean Square Tracking Errors} 
\centering
\begin{tabular}{|c||c|c|c|c|}
\hline
\hline
\multirow{2}{*}{Stiffness Delay}& \multicolumn{2}{|c|}{Experiment} & \multicolumn{2}{|c|}{Simulation}\\ \cline{2-5}
&Position & Velocity & Position &Velocity \\ 
$T_s$ (ms) & Err (rad) &  Err (rad/s) &  Err (rad) &  Err (rad/s) \\ \hline \hline
2 & 0.0182 & 0.3866  & 0.0204 & 0.3970\\ \hline
20 & 0.0247 & 0.6366 & 0.0289  & 0.6332\\  \hline 
30 & 0.0360 & 0.8753 & 0.0386  & 0.8178\\  \hline 
\end{tabular}
\label{table:RMSError}
\vspace{-0.2in}
\end{table}

\subsection{Distributed operational space control of a mobile base}
\label{subsec:doscmb}
As a concept proof of our distributed architecture on a multi-axis mobile platform, a Cartesian space feedback operational space controller (OSC) \cite{Khatib:83} is implemented on an omnidirectional mobile base. The original feedback controller can be found in \cite{Kim:13} which was implemented as a centralized process with no distributed topology at the time. The mobile base is equipped with a centralized PC computer running Linux with the RTAI realtime kernel. The PC connects with three actuator processors embedded next to the wheel drivetrains via EtherCat serial communications. The embedded processors do not talk to each other. The high level centralize PC has a roundtrip latency to the actuators of 7ms due to process and bus communications, while the low level embedded processors have a servo rate of 0.5ms. Notice that 7ms is considered too slow for stiff feedback control. To accentuate even further the effect of feedback delay on the centralized PC, an additional 15ms delay is artificially introduced by using a data buffer. Thus, the high level controller has a total of $22$ ms feedback delay.

\begin{figure*}
 \centering
    \includegraphics[width=\linewidth]{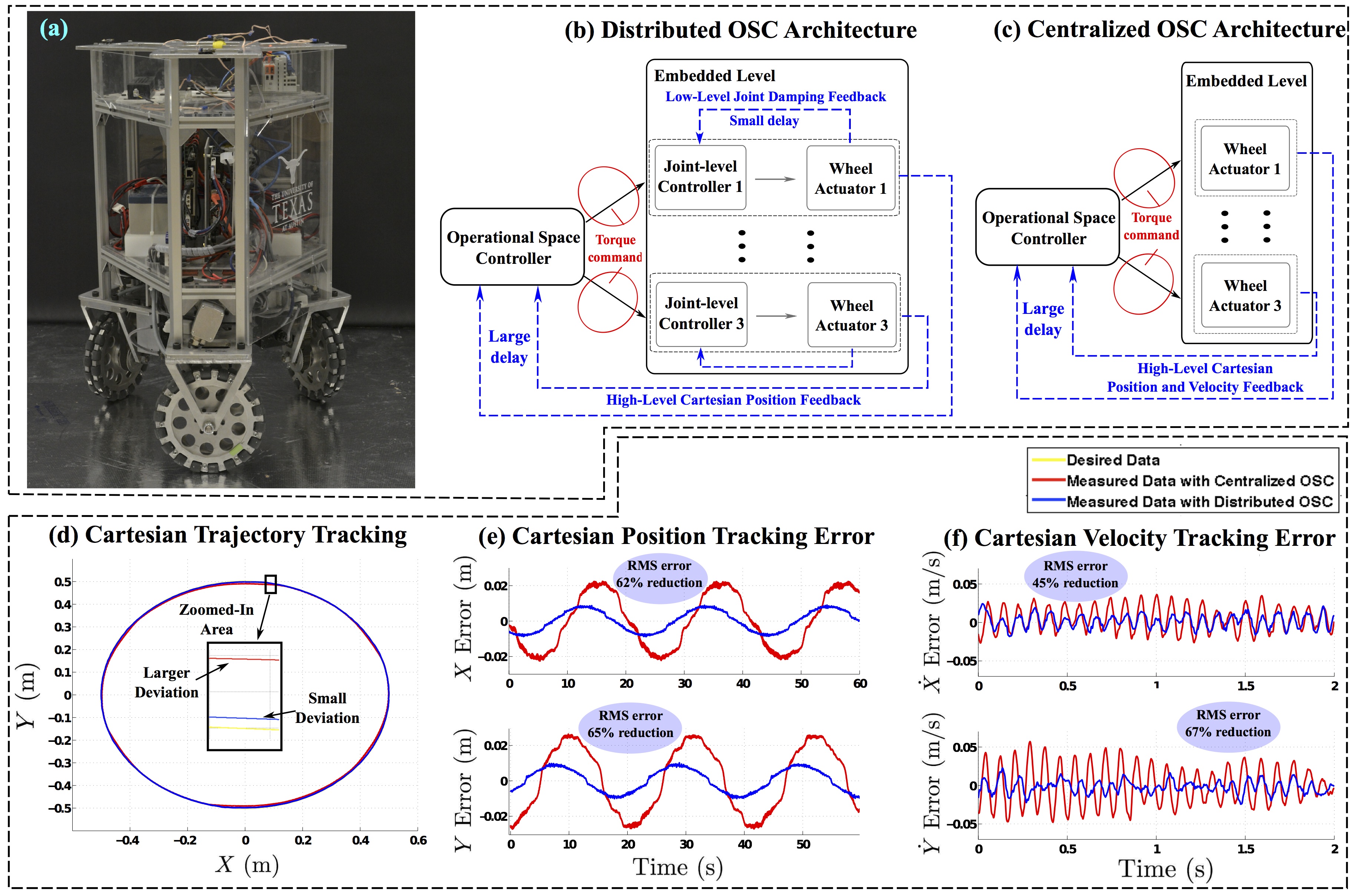}
 \caption{\textbf{Omnidirectional Mobile Base with and without Distributed Controller.} In this experiment, we implement an operational space controller based on the work published in \cite{Kim:13}. In that study, a Cartesian impedance controller is implemented in a monolithic centralized processor (see subfigure (c) for the schematic) limiting its high impedance capabilities. We re-implement that controller twice, first based on its original centralized form, COSC, and second based on the proposed distributed control architecture, DOSC, (see subfigure (b) for the schematic). The main difference between the two is that DOSC eliminates Cartesian damping feedback at the high level and instead implements joint space damping feedback at the embedded level. Due to the faster speeds of the embedded level, joint space damping feedback on the DOSC architecture can support higher impedances that are achievable with a COSC architecture. Subfigure (d) compares the Cartesian performance when the mobile base tracks a circle on the floor. Subfigure (e) demonstrates that the Cartesian position tracking errors are significantly smaller for the DOSC architecture compared to COSC. Subfigure (f) demonstrates that the Cartesian velocity error is also smaller on DOSC compared to COSC. In all experiments wheel odometry is used for trajectory tracking and error calculations based on the work published in \cite{Kim:13}.}
 \label{fig:Trikey}
\end{figure*}

An operational space controller (OSC) is implemented in the mobile base using two different architectures. First, the controller is implemented as a centralized process, which will be called COSC, with all feedback processes taking place in the slow centralized processor and none in the embedded processors. This implementation is the same as it was done for our previous work in \cite{Kim:13}. In this case, the maximum stiffness gains should be severely limited due to the effect of the large latencies. Second a distributed controller architecture is implemented inspired by the one proposed in Figure~\ref{fig:siso-controller} but adapted to our desired operational space controller, which will be called DOSC. In this version, the Cartesian stiffness feedback servo is implemented in the centralized PC in the same way than in COSC, but the Cartesian damping feedback servo is removed from the centralized process. Instead, our study implements damping feedback in joint space (i.e. proportional to the wheel velocities) on the embedded processors. A conceptual drawing of these architectures is shown in Figure~\ref{fig:Trikey}. The metric used for performance comparison is based on the maximum achievable Cartesian stiffness feedback gains, the Cartesian position tracking error, and the Cartesian velocity tracking error. To implement the Cartesian stiffness feedback processes in both architectures, the Cartesian positions and orientations of the mobile base on the ground are computed using wheel odometry and according to the methods discussed in \cite{Kim:13}. 

To achieve the highest stable stiffness gains, the following procedure is followed: (1) first, Cartesian stiffness gains are adjusted to zero while the damping gains in either Cartesian space (COSC) or joint space (DOSC) \-- depending on the controller architecture \-- are increased until the base starts vibrating; (2) the Cartesian stiffness gains, on either architecture, are increased until the base starts vibrating or oscillating; (3) a desired Cartesian circular trajectory is commanded to the base and the position and velocity tracking performance are recorded. 

Based on these experiments, DOSC was able to attain a maximum Cartesian stiffness gain of 140$\,$N/(m $\,$kg) compared to 30$\,$N/(m$\,$kg) for COSC. This result means that the proposed distributed control architecture allowed the Cartesian feedback process to increase the stiffness gains by 4.7 times with respect to the centralized controller implementation. In terms of tracking performance, the results are shown in Figure~{\ref{fig:Trikey}}. Both Cartesian position and velocity tracking in DOSC are significantly more accurate. The proposed distributed architecture reduces Cartesian position root mean error between 62\% and 65\% while the Cartesian velocity root mean error decreases between 45\% and 67\%.

\section{Conclusions and Discussion}
\label{sec:conclusion}
The motivation for this paper has been to study the stability and performance of distributed controllers where stiffness and damping servos are implemented in distinct processors. These types of controllers will become important as computation and communications become increasingly more complex in human-centered robotic systems. The focus has been first on studying the physical performance of a simple distributed controller. Simplifying the controller allows us to explore the physical effects of time delays in greater detail. Then the proposed architecture has been leveraged to a mobile base system as a proof of concept.
Our focus on this paper has been on high impedance behaviors. This focus contrasts with our previous work on low impedance control \cite{painedesign2014}. However, both high and low impedance behaviors are important in human-centered robotics. For instance, high impedance behaviors are important to attain good position tracking in the presence of unmodelled actuator dynamics or external disturbances.

Using the phase margin frequency technique allowed us to reveal the severe effects of delays on the damping loop and appreciate the discrepancy with respect to the stiffness servo behavior. However, to reveal the physical reasons for this discrepancy, in-depth mathematical analysis is performed based on phase margin sensitivity to time delays. This analysis allowed us to derive the physical condition for the discrepancy between delays. Further analysis revealed that the previous condition is met for high impedance controllers, in which the damping feedback gain is significantly larger than the passive damping actuator value. To confirm the observations and analytical derivations, three experiments are performed by using an actuator and a mobile base. In particular, the results have shown that decoupling stiffness servos to slower centralized processes does not significantly decrease system stability. As such, stiffness servo can be used to implement operational space controllers which require centralized information such as robot models and external sensors.

The next step is to develop a similar study for controllers with an inner torque loop, such as those used for series elastic actuators \cite{painedesign2014}. For this kind of actuators our interest lies in exploring both high and low impedance capabilities under latencies and using distributed control concepts similar to those explored in this paper.

\begin{appendices}

\section{Derivation of Equations~(\ref{eq:sens1}) and (\ref{eq:sens2})}
\label{sec:appendix1}

\noindent Using the product differentiation rule on Equation~(\ref{eq:sens1}), one obtains,
\begin{equation}\label{eq:app1}
\frac{\partial {\rm atan} \left(A_{1g} / A_{2g}\right) }{\partial T_s}
= \frac{\partial {\rm atan} \left(A_{1g} / A_{2g}\right) }{\partial \left(A_{1g} / A_{2g}\right) } \cdot \frac{\partial \left(A_{1g} / A_{2g}\right)}{\partial T_s}.
\end{equation}
Using differentiation rules, the following applies,
\begin{equation}
\frac{\partial {\rm atan} \left(A_{1g} / A_{2g}\right) }{\partial \left(A_{1g} / A_{2g}\right) } 
= \frac{1}{1+\left(A_{1g} / A_{2g}\right)^2}.
\end{equation}
Also, it is straightforward to derive, once again, using the product differentiation rule, the following equalities,
\begin{align}\nonumber
&\frac{\partial \left(A_{1g} / A_{2g}\right)}{\partial T_s}
= \frac{ \partial{ A_{1g} } }{\partial T_s} \frac{1}{ A_{2g} } 
+ \frac{ \partial \left(1/A_{2g}\right) }{\partial T_s} A_{1g}\\\nonumber
&\quad= \frac{ 
\left[
\begin{aligned}
&-A_{2g}\,\Big(K\omega_g\,{\rm cos}(T_s\,\omega_g) + K \, \tau_v \,\omega_g^2 \, {\rm sin}(T_s \, \omega_g)\Big)\\\nonumber
&+A_{1g} \, \Big(K\omega_g\,{\rm sin}(T_s\,\omega_g) - K \, \tau_v \, \omega_g^2 \, {\rm cos}(T_s \, \omega_g)\Big)\nonumber
\end{aligned}
\right]
}{A_{2g}^2}\\[2mm]\nonumber
&\quad=
\frac{
\left[
\begin{aligned}
&-K^2 (\tau_v^2 \omega_g^2 + 1) + KB\omega_g\,\Big({\rm sin}\left((T_s - T_d) \, \omega_g\right) \\\nonumber
&- \tau_v \omega_g {\rm cos}\left((T_s - T_d) \, \omega_g\right)\Big)\nonumber
\end{aligned}
\right]\, \omega_g
}
{A_{2g}^2},
\end{align}
which make use of the trigonometric rules ${\rm cos}^2(x) + {\rm sin}^2(x) =1$ and ${\rm sin}(x-y)={\rm sin}(x)\,{\rm cos}(y)-{\rm cos}(x)\,{\rm sin}(y)$. Gathering the above derivations and putting them together according to Equation~(\ref{eq:app1}) yields,
\begin{equation}
\frac{\partial {\rm atan} \left(A_{1g} / A_{2g}\right) }{\partial T_s}
=\quad\frac{
\left[
\begin{aligned}
&-K^2 (\tau_v^2 \omega_g^2 + 1) + KB\omega_g\,M\nonumber
\end{aligned}
\right] \, \omega_g
}
{A_{1g}^2+A_{2g}^2}
\end{equation}
with $M \triangleq {\rm sin}\Big((T_s-T_d) \;\omega_g\Big) - \tau_v \omega_g {\rm cos}\Big((T_s - T_d) \, \omega_g\Big)$. Using basic trigonometry, one can further simplify $M$ to
\begin{equation}\label{eq:MM}
M = \sqrt{\tau_v^2 \omega_g^2 + 1} \cdot {\rm sin}\Big((T_s-T_d) \;\omega_g + \phi \Big)
\end{equation}
where the phase shift $\phi \triangleq {\rm atan}(-\tau_v \omega_g)$. The derivation of Equation~(\ref{eq:sens2}) is very similar to the above and is therefore omitted for space purpose. \hspace{1in} \qed

\vspace{-0.1in}
\section{Derivation of Equation~(\ref{eq:fourthorderequation})}
\label{sec:appendix2}

\noindent Using the gain crossover frequency of Equation~(\ref{eq:unit-gain}) on the numerator of Equation~(\ref{eq:P_OLomega}) yields,
\begin{equation}
|jA_{1g}+A_{2g}|^2 = A_{1g}^2+A_{2g}^2.
\end{equation}
Using the expressions of $A_{1g}$ and $A_{2g}$ given in Equation~(\ref{eq:As}) yields,
\begin{align}\nonumber
A_{1g}^2+A_{2g}^2 &= (B\omega_g )^2 + K^2 (\tau_v^2 \omega_g^2 + 1) - 2 KB \omega_gM,
\end{align}
which also makes use of the trigonometric rules ${\rm cos}^2(x) + {\rm sin}^2(x) =1$ and ${\rm sin}(x-y)={\rm sin}(x)\,{\rm cos}(y)-{\rm cos}(x)\,{\rm sin}(y)$. Combining the above norm of the numerator of Equation~(\ref{eq:P_OLomega}) with the norm of its denominator one gets
\begin{equation}
|P_{OL}(\omega_g)|^2 = \frac{(B\omega_g )^2 + K^2 (\tau_v^2 \omega_g^2 + 1) - 2 KB \omega_g \,M}{\omega_g^2 \Big((m \omega_g)^2 + b^2\Big) \Big( (\tau_v \omega_g)^2 + 1 \Big)} = 1,
\end{equation}
which is equivalent to Equation~(\ref{eq:fourthorderequation}). \hspace{1.1in} $\qed$

\end{appendices}



\bibliographystyle{IEEEtran}
\bibliography{bib}

\begin{thebibliography}{10}
\providecommand{\url}[1]{#1}
\csname url@samestyle\endcsname
\providecommand{\newblock}{\relax}
\providecommand{\bibinfo}[2]{#2}
\providecommand{\BIBentrySTDinterwordspacing}{\spaceskip=0pt\relax}
\providecommand{\BIBentryALTinterwordstretchfactor}{4}
\providecommand{\BIBentryALTinterwordspacing}{\spaceskip=\fontdimen2\font plus
\BIBentryALTinterwordstretchfactor\fontdimen3\font minus
  \fontdimen4\font\relax}
\providecommand{\BIBforeignlanguage}[2]{{%
\expandafter\ifx\csname l@#1\endcsname\relax
\typeout{** WARNING: IEEEtran.bst: No hyphenation pattern has been}%
\typeout{** loaded for the language `#1'. Using the pattern for}%
\typeout{** the default language instead.}%
\else
\language=\csname l@#1\endcsname
\fi
#2}}
\providecommand{\BIBdecl}{\relax}
\BIBdecl

\bibitem{sakagami2002intelligent}
Y.~Sakagami, R.~Watanabe, C.~Aoyama, S.~Matsunaga, N.~Higaki, and K.~Fujimura,
  ``The intelligent asimo: System overview and integration,'' in
  \emph{Intelligent Robots and Systems, IEEE/RSJ International Conference on},
  2002.

\bibitem{diftler2011robonaut}
M.~A. Diftler, J.~Mehling, M.~E. Abdallah, N.~A. Radford, L.~B. Bridgwater,
  A.~M. Sanders, R.~S. Askew, D.~M. Linn, J.~D. Yamokoski, F.~Permenter
  \emph{et~al.}, ``Robonaut 2-the first humanoid robot in space,'' in
  \emph{Robotics and Automation, IEEE International Conference on}, 2011.

\bibitem{okamura2004methods}
A.~M. Okamura, ``Methods for haptic feedback in teleoperated robot-assisted
  surgery,'' \emph{Industrial Robot: An International Journal}, vol.~31, no.~6,
  pp. 499--508, 2004.

\bibitem{kim2005system}
J.-Y. Kim, I.-W. Park, J.~Lee, M.-S. Kim, B.-K. Cho, and J.-H. Oh, ``System
  design and dynamic walking of humanoid robot khr-2,'' in \emph{Robotics and
  Automation, Proceedings of the IEEE International Conference on}, 2005.

\bibitem{santos2006design}
V.~M. Santos and F.~M. Silva, ``Design and low-level control of a humanoid
  robot using a distributed architecture approach,'' \emph{Journal of Vibration
  and Control}, vol.~12, no.~12, pp. 1431--1456, 2006.

\bibitem{lu2014performance}
L.~Lu and B.~Yao, ``A performance oriented multi-loop constrained adaptive
  robust tracking control of one-degree-of-freedom mechanical systems: Theory
  and experiments,'' \emph{Automatica}, vol.~50, no.~4, pp. 1143--1150, 2014.

\bibitem{martin1981continuous}
J.~C. Martin and L.~George, ``Continuous state feedback guaranteeing uniform
  ultimate boundedness for uncertain dynamic systems,'' \emph{Automatic
  Control, IEEE Transactions on}, vol.~26, no.~5, p. 1139, 1981.

\bibitem{gutman1979uncertain}
S.~Gutman, ``Uncertain dynamical systems--a lyapunov min-max approach,''
  \emph{Automatic Control, IEEE Transactions on}, vol.~24, no.~3, pp. 437--443,
  1979.

\bibitem{Khatib:87(2)}
O.~Khatib, ``{A Unified Approach to Motion and Force Control of Robot
  Manipulators: The Operational Space Formulation},'' \emph{International
  Journal of Robotics and Automation}, vol. RA--3, no.~1, pp. 43--53, February
  1987.

\bibitem{wang2011decentralized}
C.~Wang and D.~Li, ``Decentralized pid controllers based on probabilistic
  robustness,'' \emph{Journal of Dynamic Systems, Measurement, and Control},
  vol. 133, no.~6, p. 061015, 2011.

\bibitem{ogata2010modern}
K.~Ogata and Y.~Yang, \emph{Modern control engineering}.\hskip 1em plus 0.5em
  minus 0.4em\relax Prentice-Hall Englewood Cliffs, 2010.

\bibitem{lee-pid}
C.-H. Lee, ``A survey of {PID} controller design based on gain and phase
  margins,'' \emph{International Journal of Computational Cognition}, vol.~2,
  pp. 63--100, 2004.

\bibitem{astrom-pid}
K.~J. \r{A}str\"{o}m, ``Automatic tuning and adaptation for {PID} controllers -
  {A} survey,'' \emph{Control Eng. Practice}, vol.~1, pp. 699--714, 1993.

\bibitem{poulin-pid}
E.~Poulin, A.~Pomerleau, A.~Desbiens, and D.~Hodouin, ``Development and
  evaluation of an auto-tuning and adaptive {PID} controller,''
  \emph{Automatica}, vol.~32, no.~1, pp. 71--82, 1996.

\bibitem{yaniv-good}
O.~Yaniv and M.~Nagurka, ``Design of {PID} controllers satisfying gain margin
  and sensitivity constraints on a set of plants,'' \emph{Automatica}, vol.~40,
  no.~1, pp. 111--116, 2004.

\bibitem{tipsuwan2004gain}
Y.~Tipsuwan and M.-Y. Chow, ``Gain scheduler middleware: a methodology to
  enable existing controllers for networked control and teleoperation-part {I}:
  networked control,'' \emph{Industrial Electronics, IEEE Transactions on},
  vol.~51, no.~6, pp. 1218--1227, 2004.

\bibitem{LeeJC14}
J.~Y. Lee, M.~Jin, and P.~H. Chang, ``Variable {PID} gain tuning method using
  backstepping control with time-delay estimation and nonlinear damping,''
  \emph{Industrial Electronics, IEEE Transactions on}, vol.~61, no.~12, pp.
  6975--6985, 2014.

\bibitem{colgate-max}
J.~Colgate and G.~Schenkel, ``Passivity of a class of sampled-data systems:
  {A}pplication to haptic interfaces,'' in \emph{American Control Conference},
  June 1994.

\bibitem{gao2008new}
H.~Gao, T.~Chen, and J.~Lam, ``A new delay system approach to network-based
  control,'' \emph{Automatica}, vol.~44, no.~1, pp. 39--52, 2008.

\bibitem{PangZH14}
Z.-H. Pang, G.-P. Liu, D.~Zhou, and M.~Chen, ``Output tracking control for
  networked systems a model based prediction approach,'' \emph{Industrial
  Electronics, IEEE Transactions on}, vol.~61, no.~9, pp. 4867--4877, 2014.

\bibitem{lawrence-impedance}
D.~Lawrence, ``Impedance control stability properties in common
  implementations,'' in \emph{Robotics and Automation, Proceedings., IEEE
  International Conference on}, vol.~2, Apr 1988, pp. 1185--1190.

\bibitem{colgate-zwidth}
J.~Colgate and J.~Brown, ``Factors affecting the {Z-Width} of a haptic
  display,'' in \emph{Robotics and Automation, Proceedings., IEEE International
  Conference on}, May 1994.

\bibitem{hulin2013optimal}
T.~Hulin, R.~G. Camarero, and A.~Albu-Schaffer, ``Optimal control for haptic
  rendering: Fast energy dissipation and minimum overshoot,'' in
  \emph{Intelligent Robots and Systems, IEEE/RSJ International Conference on},
  2013.

\bibitem{diolaiti-stability}
N.~Diolaiti, G.~Niemeyer, F.~Barbagli, and J.~Salisbury, ``Stability of haptic
  rendering: Discretization, quantization, time delay, and coulomb effects,''
  \emph{Robotics, IEEE Transactions on}, vol.~22, no.~2, pp. 256--268, April
  2006.

\bibitem{paine2014closed}
N.~A. Paine and L.~Sentis, ``A closed-form solution for selecting maximum
  critically damped actuator impedance parameters,'' \emph{Journal of Dynamic
  Systems, Measurement, and Control}, 2014.

\bibitem{paine2014actuator}
N.~Paine, J.~Mehling, J.~Holley, N.~Radford, G.~Johnson, C.~Fok, and L.~Sentis,
  ``Actuator control for the {NASA-JSC} valkyrie humanoid robot: A decoupled
  dynamics approach for torque control of series elastic robots,''
  \emph{Journal of Field Robotics}, 2014.

\bibitem{ogata-modern}
K.~Ogata, \emph{Modern Control Engineering}.\hskip 1em plus 0.5em minus
  0.4em\relax Englewood Cliffs, NJ: Prentice-Hall, 1990.

\bibitem{painedesign2014}
N.~Paine, S.~Oh, and L.~Sentis, ``Design and control considerations for
  high-performance series elastic actuators,'' \emph{IEEE/ASME Transactions on
  Mechatronics}, vol.~19, no.~3, pp. 1080--1091, 2014.

\bibitem{zhao2012three}
Y.~Zhao and L.~Sentis, ``A three dimensional foot placement planner for
  locomotion in very rough terrains,'' in \emph{Humanoid Robots, IEEE-RAS
  International Conference on}, 2012.

\bibitem{Khatib:83}
O.~Khatib, ``Dynamic control of manipulators in operational space,'' in
  \emph{IFToMM Symposium}, New Delhi, India, December 1983.

\bibitem{Kim:13}
K.~Kim, A.~Kwok, and L.~Sentis, ``Contact sensing and mobility rough and
  cluttered environments,'' in \emph{European Conference on Mobile Robots},
  September 2013.

\end{thebibliography}

\end{document}